\documentclass[twocolumn,tighten]{aastex63}
\usepackage{amssymb, amsmath,framed}

\usepackage{bm}
\expandafter\ifx\csname package@font\endcsname\relax\else
 \expandafter\expandafter
 \expandafter\usepackage
 \expandafter\expandafter
 \expandafter{\csname package@font\endcsname}
\fi
\def\bq{\begin{equation}}
\def\eq{\end{equation}}
\def\bqy{\begin{eqnarray}}
\def\eqy{\end{eqnarray}}

\def\p{\partial}

\def\p{\partial}

\def\R{\mathbb{R}}

\def\cale{\mathcal{E}}

\begin{document}
\title{\large{A Precursor Balloon Mission for Venusian Astrobiology}}

\correspondingauthor{Manasvi Lingam}
\email{mlingam@fit.edu}

\author{Andreas M. Hein}
\affiliation{Universit\'{e} Paris-Saclay, CentraleSup\'{e}lec, Laboratoire Genie Industriel, 3 rue Joliot-Curie 91190, Gif-sur-Yvette, France}
\affiliation{Initiative for Interstellar Studies (i4is), 27/29 South Lambeth Road, London, SW8 1SZ, UK}

\author{Manasvi Lingam}
\affiliation{Department of Aerospace, Physics and Space Sciences, Florida Institute of Technology, 150 W. University Blvd, Melbourne, FL 32901, USA}
\affiliation{Institute for Theory and Computation, Harvard University, 60 Garden St, Cambridge, MA 02138, USA}

\author{T. Marshall Eubanks}
\affiliation{Space Initiatives Inc., 527 Burlington Ave, Palm Bay, FL 32907, USA}

\author{Adam Hibberd}
\affiliation{Initiative for Interstellar Studies (i4is), 27/29 South Lambeth Road, London, SW8 1SZ, UK}

\author{Dan Fries}
\affiliation{Department of Aerospace Engineering and Engineering Mechanics, The University of Texas at Austin, 2617 Wichita St C0600, Austin, TX 78712, USA}
\affiliation{Initiative for Interstellar Studies (i4is), 27/29 South Lambeth Road, London, SW8 1SZ, UK}

\author{William Paul Blase}
\affiliation{Space Initiatives Inc., 527 Burlington Ave, Palm Bay, FL 32907, USA}

\begin{abstract}
The potential detection of phosphine in the atmosphere of Venus has reignited interest in the possibility of life aloft in this environment. If the cloud decks of Venus are indeed an abode of life, it should reside in the ``habitable zone'' between $\sim 50-60$ km altitude, roughly coincident with the middle cloud deck, where the temperature and pressure (but not the atmospheric composition) are similar to conditions at the Earth's surface. We map out a precursor astrobiological mission to search for such putative lifeforms \emph{in situ} with instrument balloons, which could be delivered to Venus via launch opportunities in 2022-2023. This mission would collect aerosol and dust samples by means of small balloons floating in the Venusian cloud deck and directly scrutinize whether they include any apparent biological materials and, if so, their shapes, sizes, and motility. Our balloon mission would also be equipped with a miniature mass spectrometer that should permit the detection of complex organic molecules. The mission is augmented by contextual cameras to search for macroscopic signatures of life in the Venusian atmospheric habitable zone. Finally, mass and power constraints permitting, radio interferometric determinations of the motion of the balloons in Venusian winds, together with \emph{in situ} temperature and pressure measurements, will provide valuable insights into the poorly understood meteorology of the middle cloud region.\\
\end{abstract}

\section{Introduction} \label{SecIntro}
As the surface temperature of Venus is over $460$ $^\circ$C, it is too extreme to permit the existence of life-as-we-know-it \citep{LL21}. However, if one considers the lower cloud layer of Venus at a height of $\sim 50$ km above the surface, both the temperature ($\sim 60$ $^\circ$C) and pressure ($\sim 1$ atm) become relatively clement \citep{LMS18}. This crucial fact led to pioneering proposals in the 1960s that clouds of Venus might be capable of harboring life \citep{Sag61,MS67}.

Subsequently, many studies have been undertaken to assess the plausibility of an aerial biosphere on Venus. Data from observations revealed the presence of an unknown ultraviolet (UV) absorber, which was speculated to be emblematic of biological activity \citep{Grin97,GB07}. This region exhibits several promising features including sulfur aerosols for UV protection, availability of liquid sulfuric acid and nutrients, and shielding against high-energy particles \citep{SGA04,DNP15,LMS18,SPG20}. Yet, in contrast, there are significant challenges that confront hypothetical Venusian biota, such as the extremely low water activity and the highly acidic environment due to the presence of sulfuric acid \citep{Co99,ZLM12}.

The field of Venusian astrobiology came to the forefront in $2020$ September thanks to the possible detection of phosphine at a concentration of $\sim 20$ ppb in the Venusian cloud decks by \citet{GRB20}; see, however, \citet{VCI20}. This ostensible discovery is considered significant because phosphine production on Earth almost exclusively entails biological or anthropogenic processes, and it is therefore viewed as a biosignature gas \citep{SSS20}. It has been argued that the observed abundance of phosphine could not be explained by any \emph{known} abiotic process \citep{BPS20}, implying that either novel (and unknown) abiotic mechanisms or metabolic pathways are involved. However, distinguishing between these two hypotheses is difficult, even assuming that additional data is collected and sophisticated models are developed. 

The most unambiguous method for resolving the issue of whether life exists in the Venusian atmosphere is to send spacecraft to Venus and carry out on-site measurements and experiments amid its cloud layers. This point was duly appreciated by \citet{GRB20}, who stated that: ``Ultimately, a solution could come from revisiting Venus for in situ measurements or aerosol return.'' The question that immediately springs up in this case is: what types of biomarkers are most suggestive of life? Naturally, this issue has attracted much debate, and many different classes of biosignatures have been expounded \citep{SAM08,CHP19}.

We describe two broad examples because they pertain to this work. First, in the context of organic molecules, prospective biomarkers include polymers derived from amino acids or nucleotides, enantiomeric excess of chiral amino acids and sugars, and overabundance of high-mass amino acids and other biochemical building blocks \citep{NAD20}. Second, at the level of organisms, the manifestation of cell-like morphological structures, motility, and biofabrics are believed to constitute promising indicators of life \citep{NLD16}. 

In this Letter, we map out a precursor astrobiological mission to search for putative biosignatures \emph{in situ} with a fleet of instrument balloons that could survey the Venusian atmosphere as early as 2022–2023. The outline of this Letter is as follows. We discuss the science objectives in Section \ref{SecSciO}, followed by a description of the mission architecture and stages in Sections \ref{SecArch} and \ref{SecMiss}, respectively. We summarize our findings in Section \ref{SecDisc}, while the Appendices provide the technical information.

\section{Science objectives and instruments}\label{SecSciO}
The chief science objectives of the mission are described as follows.

\begin{enumerate}
    \item \underline{Objective 1:} Collect aerosol and dust samples, and determine the shape, size, and motility of putative microorganisms (if they exist).
    \item \underline{Objective 2:} Search for macroscopic signs of life, potentially analogous to the Jovian lifeforms envisioned by \citet{SS76}.
    \item \underline{Objective 3:} Search for complex organic compounds, especially polymers composed of amino acids, nucleotides, and repeating charges. 
    \item \underline{Objective 4 (optional):} To better understand the meteorological dynamics, including zonal winds and ground-linked gravity waves, of the possible Venusian habitable zone that represents the target. Measurement of the water activity in this region could also be undertaken using a hygrometer.
\end{enumerate}

The instrumentation for the science objectives comprises the following elements.

\begin{enumerate}
   \item A combined collection plate (for gathering aerosols and dust particles) and mini-microscope with an accompanying light source (Objective 1).
   \item A camera to take contextual images of the Venusian atmosphere (Objective 2).
   \item A miniature mass spectrometer (MS) equipped with a separation stage that permits the identification of complex organics (Objective 3). In principle, the MS would also permit the characterization of inorganic compounds in the atmosphere arising from either abiotic processes or potentially metabolic pathways.
   \item Very Long Baseline Interferometry observations of
    radio emissions of each balloon together with appropriate
    meteorological instruments (Objective 4). As this objective is not
    geared toward the search for life, and as the Vega balloon mission
    conducted similar observations \citep{SKK92}, these
    measurements are not described here.
\end{enumerate}

We will begin by tackling Objectives 1 and 2. \citet{GBC11} presented a fluorescence microscope with a resolution down to $1.5$ $\mu$m and a mass of $1.9$ g. A microscope with this resolution might be sufficient to detect microbes. Another example is the Miniscope V4, with a mass of $2.6$ g and a resolution in the $\mu$m range. Other open source designs such as FinchScope, UCLA Scope, and CHEndoscope exist with masses between $1.8$ and $4.5$ g \citep{AH19}. In tandem, miniaturized collection plates and petri dishes have been delineated in the literature. For example, \citet{ISB} described a $36 \times 8$ mm petri dish on a chip with micron-scale compartments. One caveat is that microscopy techniques are usually sensitive to environmental fluctuations \citep{NLD16}, and this drawback needs to be properly investigated and addressed.

Gram-scale and even sub-gram-scale miniaturized cameras are available off the shelf for various applications. An example for a small-scale, flight-proven camera is the borescope camera on board the Visual Inspection Poseable Invertebrate Robot (VIPIR) with a diameter of $1.2$ mm and a $224 \times 224$ pixel resolution and an integrated light source for illumination.\footnote{\url{https://nexis.gsfc.nasa.gov/rrm_phase2vipir.html}} A megapixel camera of mass $\sim 0.25$ kg was employed by the Mars \emph{Curiosity} rover,\footnote{\url{https://mars.nasa.gov/msl/spacecraft/rover/cameras/}} but much smaller cameras are realizable. We estimate that the total mass of scientific instrumentation for the first two objectives can be kept to $\sim 0.1$ kg, and that they are operable at average power levels $\lesssim 1$ W after taking the power requirements of the individual components into account.

In contrast to Objectives 1 and 2, Objective 3 entails a larger payload. For instance, the miniature MS described in \citet{YKH08} has a mass of $\sim 1.5$ kg and a power consumption of $5$ W, but it is mostly suitable for detecting organic gases at ppm concentrations. At a higher mass of $3.2$ kg (sans batteries) and a power consumption of $35$ W, the tandem MS fabricated by \citet{GSH08} was shown to possess the capacity to detect proteins (or other macroscopic biomolecules) at ppb concentrations. Current technology has illustrated that liquid chromatography mass spectrometers of mass $5$ kg and power consumption of $3$ W are capable of identifying amino acid enantiomers and other biomolecular building blocks \citep{GK16}. Thus, when viewed collectively, it is likely that the MS would be a few kg and require a power expenditure of order $10$ W (see \citealt{SPO16}). However, further research is necessary to gauge whether current MS designs are capable of functioning in the environs of the Venusian atmosphere at the requisite efficiency.

\section{Precursor Venus balloon mission architecture}\label{SecArch}
Our objective is to identify a mission architecture, which can be developed at minimal mass and launched as quickly as possible, but is nevertheless capable of returning a significant amount of \emph{in situ} data from the Venusian cloud decks. We will elucidate the details of one such architecture, without claiming optimality. 

Concepts proposed for surveying the Venusian atmosphere span various types of balloons, paragliders, kites, fold-out wing gliders, solar powered aircraft, and airships \citep{D10}. We focus on balloons, as they have already been successfully operated within the Venusian atmosphere and have withstood extensive scrutiny \citep{BR20}. Venus balloons were previously utilized during the Vega 1/2 program with a total balloon mass of 21.5 kg \citep{SLB86}; a notable point of difference, however, is that the entry probe also contained a lander. Several Venusian balloon studies were subsequently conducted, such as the European Space Agency's European Venus Explorer concept \citep{CKI09}. Overviews of balloon investigations have been presented in \citet{D10} and \citet{BR20}. Current challenges for Venus missions are summarized in \citet{GLW18}: key obstacles include atmospheric entry with complex deployment and sufficient power generation.

\begin{figure*}
\begin{center}
\includegraphics[scale=0.2]{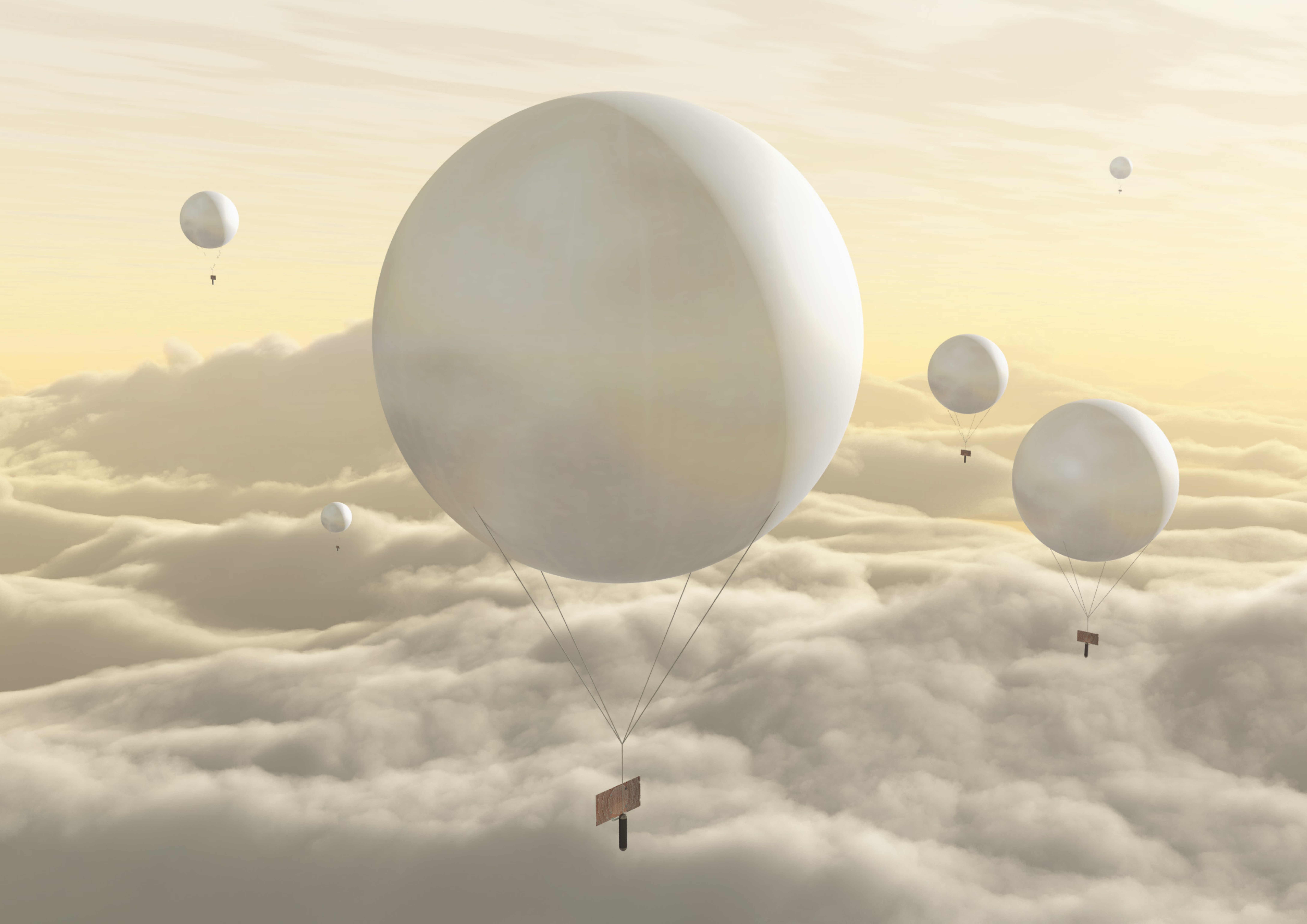}
\end{center}
\caption{Artist's impression of the balloon fleet in the upper ``clear zone'' at
$\sim 50$ km above the surface of Venus (Image credit: Adrian Mann)}
\label{fig:Ball}
\end{figure*}

Although numerous missions to analyze the Venusian atmosphere have been proposed in the past, an exclusively astrobiology-oriented mission comprising a fleet of balloons appears to be missing in the literature. We will describe two different types of balloon probes that could be put into operation in the mission, with a certain degree of redundancy built in. Figure \ref{fig:Ball} depicts the balloon fleet in the Venusian atmosphere.

\subsection{Category 1 probes}
What we label the Category 1 probes are intended to theoretically fulfill Objectives 1 and 2. These probes comprise not only the scientific payload but also a number of other key subsystems such as communications, power, harness, balloon (with fuel), and parachute. The breakdown of the various masses is delineated in Appendix \ref{SSecAppA1}, where it is determined that each probe may necessitate a total mass of $M_{t1} \sim 1.7$ kg.

There are two key points that merit further discussion. First, it was estimated in Appendix \ref{SSecAppA1} that a data rate of $\sim 10^3$ bps to Earth is not entirely unreasonable over a mission duration of $\Delta t \sim 48$ hr. In other words, over this interval, the transfer of $\sim 22$ MB is potentially realizable. If we take the camera associated with VIPIR from Section \ref{SecSciO} as an example, and employ a conversion factor of $3$ bytes per pixel, it would be possible to transmit a total of $\sim 143$ such images per each Category 1 probe. At higher resolution, however, fewer images are transmittable to Earth. This data limitation is inevitable when it comes to small spacecraft like Cubesats \citep{SK12,PG17}.

Second, we ask ourselves how many microbes would be collected over $\Delta t$. To do this, we need an estimate for the biomass number density $n_\mathrm{bio}$, which remains highly indeterminate. However, for the sake of argument, we consider $\rho_\mathrm{bio} \sim 10^{-8}$ g m$^{-3}$, which is $\sim 100$ times smaller than the biomass density near the surface of Earth \citep{FKW16}; see also \citet{LL20} for an exploration of this topic. If we choose a characteristic size of $\sim 1$ $\mu$m and a mass of $\sim 10^{-12}$ g for the microbes \citep[pg. 10]{MP16}, we obtain $n_\mathrm{bio} \sim 10^4$ m$^{-3}$. If these microbes possess a settling velocity of $\bar{v}$, then the amount of microbes $N_m$ collected over the area $A$ is $N_m \sim n_\mathrm{bio} \bar{v} A \Delta t$. We choose $\bar{v} \sim 10^{-3}$ m s$^{-1}$ based on the estimate for $1$ $\mu$m aerosols \citep[pg. 2178]{JCM61}. Note that the mass of Category 1 probes is roughly equivalent to that of a standard Cubesat with cross-sectional area of $10$ cm by $10$ cm \citep{SK12}, of which $\sim 1\%$ can be set aside for microbe collection; moreover, the resultant area of $A \sim 10^{-4}$ m$^2$ is comparable to the total area of the petri dish mentioned in Section \ref{SecSciO}. 

By substituting these fiducial values, we obtain $N_m \sim 170$ microbes. Thus, even if the biomass density is two orders of magnitude smaller than the value adopted here, it seems conceivable that each Category 1 probe might stumble across a Venusian microbe.

\subsection{Category 2 probes}
The mass of the Category 2 probe is higher relative to Category 1 probes because the former is equipped with the MS to carry out Objective 3. We invoke the tandem MS that was described in Section \ref{SecSciO} for our purposes. As with the Category 1 probes, the analysis of various subsystems is undertaken in Appendix \ref{SSecAppA2}. Based on these estimates, we calculate a mass of $M_{t2} \sim 16.1$ kg for the Category 2 probe, which is nearly an order of magnitude higher than the Category 1 probes. 

As indicated in Appendix \ref{SSecAppA2}, we have posited a data rate that is about $5$ times higher at $\sim 5 \times 10^3$ bps; note that this value is about an order of magnitude removed from the data rates associated with large-scale missions such as \emph{Cassini}.\footnote{\url{https://solarsystem.nasa.gov/news/12976/cassinis-largest-science-instrument/}} This choice of the bit rate implies that the total amount of data transmitted to Earth is $\sim 108$ MB. If we utilize the camera specifications for VIPIR, a total of $\sim 715$ images could be sent back to Earth. In this case, however, it should be appreciated that the MS data is not encapsulated by images as such.

\begin{figure*}
\begin{center}
\includegraphics[width=2.\columnwidth]{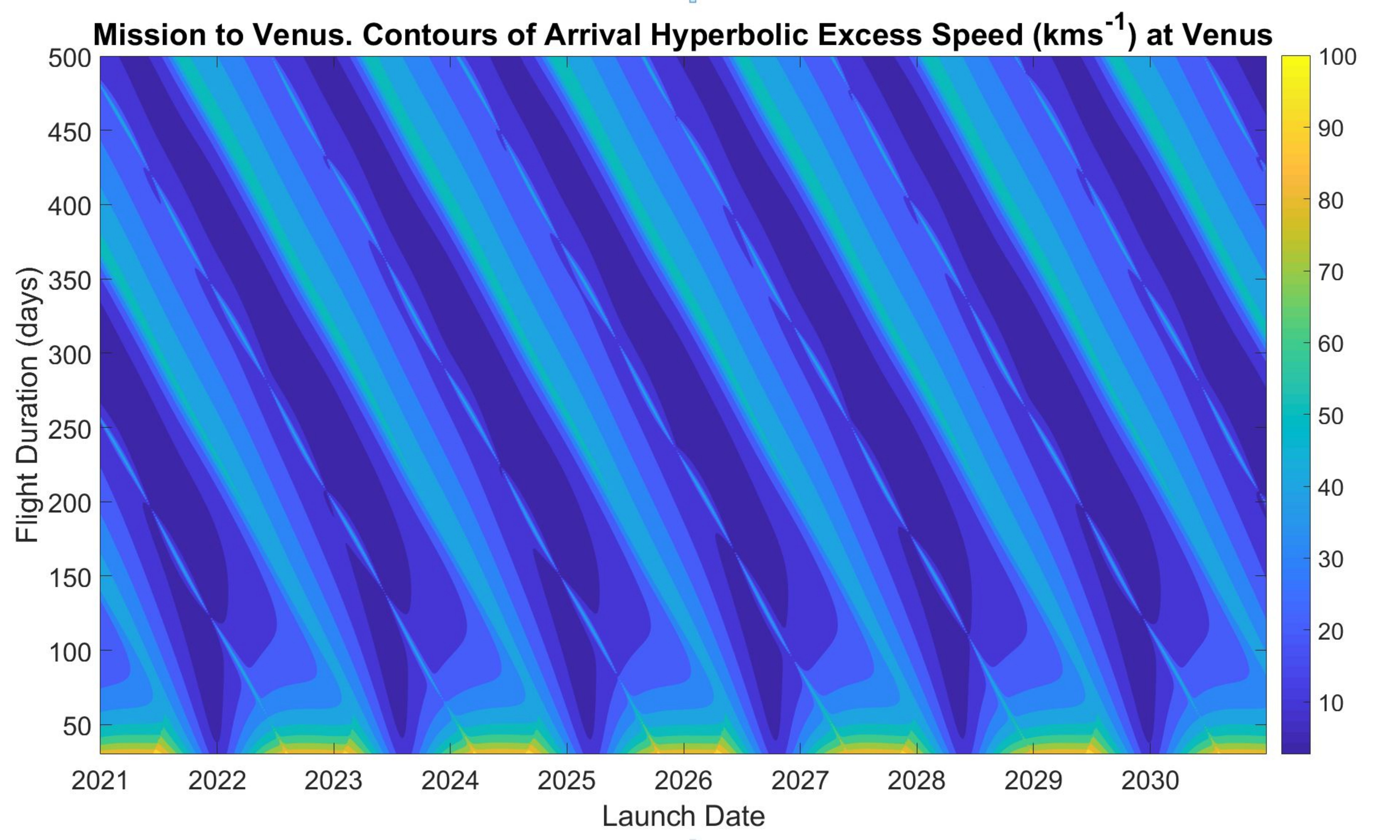}
\end{center}
\caption{Porkchop plots for encounter velocities at Venus from Earth in km s$^{-1}$. An ideal mission would have both short transfer time and low entry velocity. Such missions are possible every synodic period ($\sim 584$ days) with the next such launch opportunity arising in $2021$-$2022$.}
\label{fig:velocities}
\end{figure*}

\begin{table*}
\begin{minipage}{165mm}
\caption{Breakdown of mass requirements for various subsystems}
\label{TabSubSys}
\vspace{0.1 in}
\begin{tabular}{|c|c|c|}
\hline 
Subsystem & Mass (in kg) & Notes\tabularnewline
\hline 
\hline 
Category 2 scientific instruments & 3.2 & Section \ref{SecSciO} \tabularnewline
\hline 
Category 2 gondola & 8.3 & Appendix \ref{SSecAppA2}\tabularnewline
\hline 
Category 2 balloon mass & 4.6 & Appendix \ref{SSecAppA2}\tabularnewline
\hline 
\textbf{\textit{Total Category 2 probe mass}} & 16.1 & Summation of above three components\tabularnewline
\hline
Category 1 scientific instruments & 0.1 & Section \ref{SecSciO} \tabularnewline
\hline 
Category 1 gondola & 1.1 & Appendix \ref{SSecAppA1}\tabularnewline
\hline 
Category 1 balloon mass & 0.5 & Appendix \ref{SSecAppA1}\tabularnewline
\hline 
\textbf{\textit{Total Category 2 probe mass}} & 1.7 & Summation of above three components\tabularnewline
\hline
Total mass of all Category 2 probes & 8.5 & Five probes of this type\tabularnewline
\hline
\textbf{\textit{Total mass of all probes}} & 24.6 & Sum of Category 1 and 2 balloons\tabularnewline
\hline
\hline
Entry and descent system & 12.3 & Section \ref{SecMiss} \tabularnewline
\hline 
\textbf{Total mass entry vehicle without margin} & 36.9 & Sum of all probes and entry/descent system\tabularnewline
\hline 
\textbf{Total mass entry vehicle with 30\% margin} & 48.0 & Previous entry multiplied by a factor of 1.3 \tabularnewline
\hline 
\end{tabular}
\medskip

\textbf{\textit{Additional notes:}} The ``gondola'' is assumed to encompass the relevant subsystems such as power, communications, and harness. The ``balloon mass'' includes the mass of the fuel, balloon fabric, and parachute. Whenever a particular Section or Appendix is listed, the rationale for the mass specification is explained therein.
\end{minipage}
\end{table*}

\section{Mission stages}\label{SecMiss}
To implement the mission, it is necessary to identify a suitable launcher for transporting our entry vehicle, which is composed of the Category 1 and Category 2 probes outlined previously.

In terms of technology, we consider existing small launchers, notably Rocket Lab's Electron vehicle with an upper stage, which is theoretically capable of launching payloads of $68$ kg for a Venus flyby mission and a Venus entry probe with a mass of $37$ kg.\footnote{\url{https://www.space.com/rocket-lab-venus-life-hunting-mission.html}} The spacecraft subsystems, for the most part, should rely on off-the-shelf and commercial-off-the-shelf technologies to decrease development duration. The mission sequence consists of launch and Earth escape, Venus arrival, direct atmospheric entry and descent, balloon deployment, and continued balloon operations until the onset of balloon failure. Previous studies have relied on prior orbit insertion, as it facilitates orbiter deployment \citep{VFA06}, but we opt for direct atmospheric entry.

We have discussed balloon deployment and operations in Section \ref{SecArch} and Appendix \ref{SecAppA}. We will not address the launch and Earth escape, because these standard problems have been widely investigated in the literature. Figure \ref{fig:velocities} shows the range of Venus entry velocities as a function of launch date and flight duration. The plot was generated via the Optimum Interplanetary Trajectory Software (OITS) (\url{https://github.com/AdamHibberd/Optimum_Interplanetary_Trajectory}). It adopts the patched conic assumption, and solves Lambert’s problem for Earth departure and arrival at Venus. The resulting non-linear global optimization problem with inequality constraints is solved by applying the NOMAD solver \citep{L11}. 

The lowest perihelion velocity is achievable through a Hohmann transfer, which yields $v_\infty = 2.7$ km s$^{-1}$ at the boundary of the Venusian sphere of influence. Together with the second Venusian escape velocity and the vis-viva equation $v=\sqrt{v_\infty^2+v_{esc}^2}$, where $v_{esc}$ is the second escape velocity of Venus, one obtains $v = 10.7$ km s$^{-1}$ as the minimum velocity for direct Venus atmospheric entry from interplanetary space. This entry velocity is higher than the entry velocity of $8$ km s$^{-1}$ from the International Space Station (ISS), which necessitates the dissipation of $\sim 80\%$ higher kinetic energy during entry, and is similar to entry conditions into the Earth atmosphere from the Moon. 

We rely on proven entry technology, such as the scaled-up version of the Reentry Breakup Recorder (REBR), which is one of the smallest proven Earth re-entry capsules to date with a total mass of 4 kg \citep{WA12,FWA13}; it has been utilized for re-entry from the ISS.\footnote{\url{https://www.nasa.gov/mission_pages/station/research/news/rebr.html}} Due to similarities in the atmospheric entry conditions on Earth and Venus, the technology should be adaptable to Venus entry by incorporating a proportionally larger heat shield. If there are $n_1$ probes of Category 1 and $n_2$ probes of Category 2, the total payload mass for this entry probe is $M_t = n_1 M_{t1} + n_2 M_{t2}$. In what follows, we work with $n_2 = 1$ because of the higher mass associated with Category 2 probes, but we remark that this assumption can be easily relaxed. There is, however, a crucial missing component: heat shields. They comprise a sizable fraction of the mass of the entry vehicle. Other components, such as the balloons release mechanism, are anticipated to involve lower mass constraints.

The exact mass of the shields varies quite significantly depending on atmospheric properties, entry velocities, and many others. The ratio of the heat shields to that of the entry vehicle mass is $\sim 0.1$-$0.5$ \citep[Table 3.5]{BGLK07}. We will err on the side of caution and presume that the total mass of the entry vehicle ($M_v$), which also encompasses shields, scales linearly with $M_t$ (see \citealt{HW09}) and entails a conversion factor of $\zeta \approx 1.5$, to wit, we have $M_v \sim \zeta M_t$. If we select $M_v \approx 37$ kg, we find $M_t \sim 24.7$ kg and $n_1 \sim 5$. In contrast, if we choose a higher vehicle mass of $M_v \sim 68$ kg, we arrive at $M_t \sim 45.3$ kg and $n_1 \sim 17$.

We carry out a consistency check by selecting $M_t \sim 24.7$ kg to begin with. We consider an entry velocity range of $10.7-12$ km s$^{-1}$ and assume deceleration to rest. After including the potential energy of the vehicle, using Phenolic Impregnated Carbon Ablators (PICA) as the heat shield material with an enthalpy of ablation $0.233$ GJ kg$^{-1}$,\footnote{\url{https://ntrs.nasa.gov/citations/19970017002}} and adopting a $\sim 20\%$ safety factor for the heat shield, we determine that the mass of the heat shield required spans $11.2$ kg to $14$ kg for entry speeds between $10.7$ km s$^{-1}$ and $12$ km s$^{-1}$, respectively. If we add this heat shield mass to $M_t$, we notice that the total mass is either below or close to the stipulated vehicle mass of $M_v \sim 37$ kg as desired.

However, in undertaking the above calculations, we did not incorporate a mass margin, which is generally standard practice. If we choose an approximate mass margin of $30\%$, then the relationship linking $M_v$ and $M_t$ is transformed into $M_v \sim \kappa \zeta M_t$, where we have $\kappa \approx 1.3$. By repeating the analysis for $M_v \approx 37$ kg, we obtain $M_t \sim 19$ kg and $n_1 \sim 2$. On the other hand, for $M_v \sim 68$ kg, we end up with $M_t \sim 34.9$ kg and $n_1 \sim 11$.

Hence, there are a couple of broad takeaways from the preceding estimates. In the event that the entry vehicle mass is constrained to be $37$ kg, it is impossible to have many Category 1 probes unless one sacrifices the mass margin; in the latter case, a five-fold redundancy in the Category 1 probes would seem feasible. In contrast, if the launcher can transport a vehicle of mass $68$ kg to Venus, the mission might permit $> 10$ Category 1 probes even with a comfortable mass margin of $30\%$, thereby allowing for enhanced redundancy.

\begin{table*}
\begin{minipage}{150mm}
\caption{Breakdown of costs involved in the precursor mission}
\label{tab:Cost}
\vspace{0.1 in}
\begin{tabular}{|c|c|c|}
\hline 
System & Cost in M\$ & Notes\tabularnewline
\hline 
\hline 
Launcher cost & 10 & Baseline launcher plus upper stage cost; estimate in Appendix \ref{SecAppB} \tabularnewline
\hline 
Probe cost & 2.4-9.6 & Cost range of $\$50$-$200$ thousand kg$^{-1}$ \citep[pg. 808]{WL99}; \\
{} & {} & estimate in Appendix \ref{SecAppB} \tabularnewline
\hline 
\hline 
\textbf{\textit{Total cost}} & 12.4-19.6 & Sum of the above two costs\tabularnewline
\hline
\end{tabular}
\medskip
\end{minipage}
\end{table*}

\section{Discussion}\label{SecDisc}
The putative (albeit contested) detection of phosphine in the Venusian atmosphere has reignited interest in sending life-detection missions to our sister planet. Motivated by the rapid technological growth and versatility of small spacecraft such as Cubesats \citep{SK12,PG17}, we have delineated a possible template for precursor missions aiming to search for indicators of life in the cloud decks of Venus.

This Letter has demonstrated that a low-mass, low-cost precursor vehicle to explore the Venusian cloud layers of interest to astrobiology could be constructed and launched within the next $2-3$ yr at a budget of $< \$20$ million and mass of $\sim 40$-$50$ kg, as elucidated in Tables \ref{TabSubSys} and \ref{tab:Cost}. This cost range has the benefit of placing this mission within reach of private initiatives. Furthermore, we based our analysis on existing technologies such as the Electron launcher, off-the-shelf technologies for the various subsystems (e.g., balloons and communications), and upscaled REBR technology for the entry and descent system. We also incorporated a certain degree of redundancy in the mission by allowing for the existence of multiple probes. The life-detection mission might be able to collect data of significant astrobiological interest by way of measuring the composition of dust and aerosols via the mass spectrometer, $\mu$m-scale particles and structures via microscopes, and potential macroscopic biogenic signatures via cameras.

In closing, we stress that our proposal should be viewed as a preliminary template and forerunner for more comprehensive studies; therefore, it is not the only viable route. For instance, the mission could be scaled upward or downward in terms of mass and power, and the choice of instrumentation for the scientific payloads is also flexible because one can swap the designated instruments on some probes with others of similar mass and power without altering our conclusions. By doing so, the architecture may permit a broader spectrum of scientific objectives - extending beyond astrobiology into various domains of planetary science - to be fulfilled. Hence, future research along these lines, including in-depth subsystem-level engineering, is warranted.

\acknowledgments
We thank Cassidy Cobbs and Robert Kennedy for the insightful comments and discussions. We are grateful to our reviewer, Chris McKay, for the helpful and meticulous report. ML acknowledges the support provided by the Florida Institute of Technology.

\appendix 

\section{Masses of Category 1 and Category 2 Probes}\label{SecAppA}
Here, we describe the rationale underlying the masses of the Category 1 and Category 2 probes.

\subsection{Mass of Category 1 Probes}\label{SSecAppA1}
It was noted earlier in Section \ref{SecSciO} that the scientific instruments may entail mass and power requirements of $M_{s1} \sim 0.1$ kg and $P_{s1} \sim 1$ W, respectively. There are, however, many other subsystems that come into play, of which we shall tackle only the most salient ones.

We first examine the crucial issue of communications. In \citet[Table 10.4]{BGLK07}, the data rate received at Earth by a $32$ m radio telescope is $\sim 10$ bps, after assuming a $P_{c1} \sim 5$ W transmitter operating at $2.3$ GHz. However, in \citet{BGLK07}, a conservative transmitter gain of unity was considered. In contrast, the high-gain antenna developed in the context of the \emph{Mars Cube One} Cubesat mission was capable of reaching a gain of $\sim 10^3$ \citep{HCH17}. If we choose a lower transmitter gain of $\sim 100$ instead, we obtain a data rate of $\sim 10^3$ bps. A major advantage of the antenna described in \citet{HCH17} is that it can fit into a Cubesat, has a power requirement of $P_{c1}$, and is characterized by a total mass of $< 1$ kg. As the antenna that we have outlined has smaller gain and size footprint, we adopt a fiducial mass of $M_{c1} \sim 0.5$ kg. 

There are two additional distinct possibilities that deserve to be explicated here. In the first, the larger balloon (i.e., the Category 2 probe) acts as a relay, whereby the data is transmitted by Category 1 probes to Earth. In the second scenario, an opportunistic data link to an orbiter or flyby probe is set up. For example, India has tentatively scheduled the launch of the Shukrayaan-1 spacecraft to Venus in 2023. On the one hand, these opportunistic links could permit data rates several orders of magnitude higher than a direct link. On the other hand, a significant disadvantage that arises is the limited line of sight to the orbiter.

It might be desirable to use supercapacitors in lieu of conventional batteries, but further research is needed to substantiate this option. In comparison to lithium-ion batteries, they have higher specific power and lower specific energy \citep{horn2019supercapacitors}. Thus, they are suitable for miniaturized devices and relatively short-term missions with high, albeit intermittent, power demands. Graphene-based supercapacitors with dimensions of $10\,\mathrm{mm}\times17\,\mathrm{mm}$ and thickness of order $0.1$ mm are already available \citep{djuric2017miniature}.

Next, we turn our attention to the power required for the scientific instruments and the communications link. We have seen that a total power of $P_{t1} \sim P_{s1} + P_{c1}$ is required, based on which a total power of $P_{t1} \sim 6$ W is calculated. If we wish to deploy these two subsystems over a time $\Delta t$ (in hours), the energy expenditure is $P_{t1} \Delta t$. For a specific energy of $\cale \approx 500$ W hr kg$^{-1}$, which is smaller than state-of-the-art experimental technologies by a factor of $> 5$ \citep{KOT19}, the mass required to furnish this power for this balloon is $M_{b1} \approx P_{t1} \Delta t/\cale$. If we specify a mission duration of $\Delta t \sim 48$ h, we have $M_{b1} \sim 0.6$ kg. However, this might represent an upper bound in some respects because it assumes that all of the subsystems are continuously functional. 

The power can be extracted from solar energy instead. Solar flux values within the Venus atmosphere were summarized in \citet{TBC07}, by using data from the Pioneer Venus LSFR experiment spectrophotometers on board the Venera 11, 13, 14 landers. At an altitude of $60$ km, solar flux values range from $400$ to $1000$ W m$^{-2}$. At $50$ km, the range is between $200$ and $400$ W m$^{-2}$. For state-of-the-art solar cells with a conversion efficiency of $30 \%$, we get an area-specific power of $60$-$300$ W m$^{-2}$. A collection area of $\sim 1$ m$^2$ would therefore yield $\gtrsim 100$ W and necessitate a mass of a few kg.\footnote{\url{https://www.nasa.gov/smallsat-institute/sst-soa/power}} While the use of solar panels might be cheaper in terms of mass, their usefulness will diminish as the probes are swept away at horizontal speeds of $\sim 100$ m/s due to Venus' superrotating atmosphere. In principle, a combination of battery and solar power is probably ideal, but we will not explicate such hybrid designs in this prefatory Letter.

Thus, the total mass of the payload is $M_{p1} \sim 1.2$ kg after summation of the prior masses. The payload is connected to the balloon by means of an appropriate harness structure. The mass of the suspension system is typically negligible because it constitutes $\lesssim 10\%$ of the total payload mass for some past missions \citep[Tables 23.1 and 26.1]{BGLK07}; see also \citet{VFA06}. We will proceed with this apparently reasonable premise hereafter, given that the mission takes place in the Venusian cloud layer, whose conditions resemble those of Earth's atmosphere near our planet's surface in many respects.

We select a light gas zero-pressure balloon motivated by the simplicity of the design, with adjustments to ambient pressures implemented through vents. Although this choice limits the overall lifetime of the balloon, we have deliberately opted for a simpler design to reduce complexity. In actuality, the Venusian atmosphere is beset by a number of drawbacks such as the corrosive effects of sulfuric acid, high wind speeds and elevated pressures at lower altitudes. However, superpressure balloon prototypes have been constructed to bypass these issues, thus permitting survival on timescales of several days \citep{HFF08}, which is more stringent than our design parameters. For these balloons, the ratio of balloon mass to payload mass is approximately $0.38$ \citep{HFF08}, and this is close to the conversion factor of $\sim 0.3$ for the loaded balloons including fuel \citep[Table 6.2]{BGLK07}.

The Category 1 probes are deployed from an entry vehicle that is endowed with the requisite thermal shielding, as discussed in Section \ref{SecMiss}. However, given that the Category 1 probes will nonetheless enter at speeds of $\sim 10$ km s$^{-1}$, achieving a terminal velocity of $< 10$ m s$^{-1}$ is advisable. This reduction can be effectuated by means of a parachute. For a probe of mass $\sim 1$ kg, the diameter of the parachute can be computed by invoking \citet[Equation 4.1]{BGLK07}, which yields $\sim 1$ m.\footnote{\url{https://apogeerockets.com/education/downloads/Newsletter149.pdf}} Despite their large cross-sectional area, the mass of a parachute is very small. For instance, a parachute of $\sim 1$ m diameter might possess a mass of $\lesssim 0.1$ kg.\footnote{\url{https://www.highaltitudescience.com/products/0-9-m-parachute}} Thus, in line with the above considerations, a scaling factor of $\epsilon = 1.4$ is introduced to convert the total payload mass to the total mass $M_{t1}$ of the Category 1 probe: from this linear scaling, we obtain $M_{t1} \sim \epsilon M_{p1} \sim 1.7$ kg. 

\subsection{Mass of Category 2 probes}\label{SSecAppA2}
The mass and power of the scientific instruments is dominated by the tandem MS, which is taken to necessitate $M_{s2} \sim 3.2$ kg and $P_{s2} \sim 35$ W, respectively. Alternative designs are capable of reducing the power requirements to an extent, but may run the risk of losing the stipulated sensitivity. Even if other instruments accompanying the Category 1 probe are incorporated herein, the mass is only weakly affected. The same also applies to the inclusion of auxiliary devices such as low-mass quadcopters with masses of $\lesssim 0.1$ kg. We will now investigate the various subsystems in the same vein as Appendix \ref{SSecAppA1}.

For the communications link, we take our cue from the detail that the Category 2 probe will end up being more massive than the Category 1 probe by almost an order of magnitude owing to the heavier scientific payload. Therefore, we scale the mass and power of the communications system by a factor of $\sim 5$ relative to the Category 1 probe. In other words, by employing Appendix \ref{SSecAppA1}, we select $P_{c2} \sim 25$ W and $M_{c2} \sim 2.5$ kg; we have supposed that the transmitter power scales linearly with the mass. In this case, \emph{ceteris paribus}, the data rate is enhanced to $\sim 5 \times 10^3$ bps. 

The power required by the Category 2 probe is given by $P_{t2} \sim P_{s2} + P_{c2} \sim 60$ W. Under the assumption that the battery technology in Appendix \ref{SSecAppA1} can be scaled to higher masses, we are in a position to deploy $M_{b2} \approx P_{t2} \Delta t/\cale$. By substituting the appropriate values into this formula, we obtain $M_{b2} \sim 5.8$ kg. Note that this battery mass embodies an upper bound because all subsystems were taken to be continuously operative. 

The total payload mass for this probe consequently adds up to $M_{p2} \sim 11.5$ kg. Following the same line of reasoning described in Appendix \ref{SSecAppA1}, the mass of the harness is neglected. In order to account for the additional mass contributed by the balloon, fuel and the parachute, we utilize the factor $\epsilon$ introduced earlier. Thus, as per this scaling, the total mass $M_{t2}$ of the Category 2 probe is given by $M_{t2} \sim \epsilon M_{p2} \sim 16.1$ kg. 

\section{Programmatics: Cost and Schedule}\label{SecAppB}
We use a simple cost and schedule model to get ballpark estimates for the proposed life-detection mission. The first item that we tackle is the cost of the launcher. The cost of one Electron launch vehicle with the upper stage is about $\$ 10$ million, of which $\$ 6$ million is the baseline cost vehicle;\footnote{\url{https://www.sciencemag.org/news/2017/12/rocket-lab-poised-provide-dedicated-launcher-cubesat-science}} other sources point toward an even lower baseline cost of $\$ 5$ million.\footnote{\url{https://directory.eoportal.org/web/eoportal/satellite-missions/e/electron}} We have therefore added $\$ 4$-$5$ million as a rough estimate for the cost incurred by the upper stage and other components. 

In order to gauge the development cost for the probe, including the entry and decent vehicle and balloons, we select a specific cost of $\$50$-$200$ thousand kg$^{-1}$, which corresponds to the range of values provided in \citet[pg. 808]{WL99}. Our choice may represent a conservative selection because some of the developmental costs have decreased over time. For a mass budget of $48$ kg delineated in Table \ref{TabSubSys}, the development cost amounts to $\$ 2.4$-$9.6$ million. The final costs associated with the life-detection mission are tabulated in Table \ref{tab:Cost}. The total cost of $< \$ 10$ million ought to enable private investors and/or national agencies to finance this precursor mission.

If one assumes typical privately developed small spacecraft - such as the Electron launch vehicle mentioned above - and draws upon off-the-shelf technology, a development duration of $2$-$3$ yr appears realistic \emph{prima facie}, which would permit a launch in the $2022$-$2023$ timeframe.

\bibliographystyle{aasjournal}
\bibliography{VenusMission}

\end{document}